\documentstyle[floats,prd,aps]{revtex}
\begin{document}
\draft
\title{Disk collapse in general relativity}
\author{Andrew M. Abrahams\cite{abrahams},
Stuart L. Shapiro\cite{shapiro}, and
Saul A. Teukolsky\cite{teukolsky}}
\address{Center for Radiophysics and Space Research \\
         Cornell University
         Ithaca, NY 14853}
\date{\today}
\twocolumn[
\maketitle
\begin{abstract}
\widetext
The radial collapse of a homogeneous disk of collisionless particles
can be solved analytically in Newtonian gravitation. To solve the problem
in general relativity, however, requires the full machinery of numerical
relativity. The collapse of a disk is the
simplest problem that exhibits the two most significant and
challenging features of strong-field gravitation: black hole formation and
gravitational wave generation. We carry out dynamical calculations
of several different relativistic disk systems. We explore the growth of
ring instabilities in equilibrium disks, and how they are suppressed by
sufficient velocity dispersion. We calculate wave forms from oscillating
disks, and from disks that undergo gravitational collapse to black holes.
Studies of disk collapse to black holes should also
be useful for developing new techniques
for numerical relativity, such as apparent
horizon boundary conditions for black hole spacetimes.
\end{abstract}
]
\narrowtext

\section{Introduction}

The simplest example of gravitational collapse is that of
a homogeneous sphere of particles initially at rest.
This collapse solution is analytic both in Newtonian gravity and
general relativity.  In general relativity, this solution
is known as  Oppenheimer-Snyder collapse\cite{opp} (the
solution is a ``piece'' of a closed Friedmann universe).
Because of Birkhoff's theorem we know that this solution
is nonradiating.  Both the particle motion and the gravitational
field are radially symmetric, i.e., functions of one spatial
variable.

The radiating problem which is the simplest analogue to
Oppenheimer-Snyder collapse is that of an axisymmetric,
infinitely thin disk of particles initially at rest.  This
case is simple
because the particle motion still depends on only one spatial
coordinate, although the gravitational field now depends on
two.  If the disk is constructed by squashing a homogeneous sphere
into a pancake, keeping the density homogeneous, then the
solution is still analytic in Newtonian theory\cite{scalar}.

However, to solve the problem for relativistic gravitation
requires the full machinery of numerical relativity, and has not been addressed
until
now. Indeed, the dynamical properties of a disk of collisionless particles
has never been studied in general relativity, although it has been
extensively treated in Newtonian theory\cite{binneytremaine}.
In this paper we tackle the collapse of an axisymmetric collisionless
disk of particles in general relativity theory.  Particles in an axisymmetric
disk can also have angular motion, but because of conservation of
angular momentum this motion is not dynamical.  We consider cases in
which the disk particles are initially at rest and also in which they
have initial angular motion.  In the latter case we focus
on disks with total $J=0$, i.e. with equal numbers of co- and counter-
rotating particles.

In Newtonian theory, it is known that equilibrium disks supported
against collapse by rotation alone are unstable to ring
formation\cite{binneytremaine}.
The disk can be stabilized by ``heating" the disk, that is, converting some
of the ordered rotational energy into random ``thermal" motion  \cite{kalnajs}.
We explore here whether or not a similar result holds in general relativity.

Our study of disks is geared to analyze two relativistic phenomena
that do not arise in Newtonian theory: collapse to black holes
and the generation of gravitational waves.  Since this is the simplest
wave generation problem, disk collapse provides a useful proving
ground for testing codes designed to treat gravitational radiation
in general relativity.  Once we can evolve disk systems accurately
in general relativity, we should be able to compute the gravitational
field from any axisymmetric source, since the equations for the field
are essentially the same as those for general axisymmetric sources.
The major problems associated with general axisymmetric collapse
are all contained in this test case:  formation and
evolution of black holes and propagation of gravitational waves.

We have previously studied disk collapse in the context of nonlinear
scalar gravitation, where many of the techniques for handling
infinitely thin disks were developed\cite{scalar}.
The basic code for evolving axisymmetric spacetimes in
general relativity has been discussed in a number of papers
\cite{grcm,coll,grcmrot}.  We will refer extensively to the equations
in Ref.~\cite{coll} in the discussion that
follows.  The organization
of this paper is as follows. In Sec.\ref{sec:equations}
we present the equations for an evolving, general relativistic disk.
In Sec.~\ref{sec:analytic} we discuss analytic test problems
including Newtonian and relativistic disk solutions.
In Sec.~\ref{sec:results} we give the results of our numerical
calculations.
\section{Basic equations}\label{sec:equations}
The particles comprising the disk are assumed to interact
exclusively by gravitation, i.e., they obey the relativistic
collisionless Boltzmann equation (Vlasov equation).
Accordingly,  we have constructed a numerical code that solves
Einstein's equations for the gravitational field coupled to matter
sources obeying the Vlasov equation. This is the same mean-field,
particle simulation code described in
Refs.~\cite{grcm,coll} to study nonspherical gravitational collapse.
The code is designed to handle axisymmetric systems with no
net angular momentum. The present version assumes equatorial symmetry.
We solve the field equations in $3+1$ form following Arnowitt, Deser,
and Misner\cite{adm}.  We use maximal time slicing and quasi-isotropic
spatial gauge in axisymmetry.  The metric, written in
spherical-polar coordinates is
\begin{eqnarray}
ds^2=-\alpha^2 dt^2+A^2(dr+\beta^r dt)^2+A^2r^2(d\theta+\beta^\theta dt)^2
\nonumber \\
+B^2r^2\sin^2\theta d\phi^2.
\end{eqnarray}
The matter satisfies the relativistic Vlasov equation, which we
solve by particle simulation in the mean gravitational field.
The basic code is identical to the one described in Refs. \cite{grcm}.
The key equations and
definitions of variables are given in the Appendix of Ref. \cite{coll}.
(These were generalized to include net rotation in Ref. \cite{grcmrot}.)

\subsection{Jump conditions}

The disk matter source affects the metric via jump
conditions in the field equations across the equatorial
(disk) plane.  These jump conditions replace the usual matter
source terms that appear in the field equations.  A derivation
and numerical implementation of such a jump condition was
presented in Ref.\cite{scalar} for the simple case of scalar
gravity.   There the governing equation is a nonlinear wave equation
with a matter source on the right-hand-side.

As an example of how the jump conditions may be derived in $3+1$
general relativity, consider the two momentum constraint equations (A4) and
(A5) in Ref.\cite{coll}:
\begin{eqnarray}
{1\over{\sin \theta}}\partial_\theta(\sin \theta~ \hat K^r_{~r})
+ {T\over{\sin^2\theta}}
\partial_\theta\left({{\sin^2\theta}\over T} \hat K^\phi_{~\phi}\right)
&=& -S_\theta  \nonumber \\ +
{1\over r^2}\partial_r (r^2 \hat K^r_{~\theta}), \label{momtheta} \\
{1\over r^3}\partial_r(r^3\hat K^r_{~r}) + \hat K^\phi_{~\phi}\,\partial_r\eta
&=& S_r \nonumber \\
-{1\over{r^2 \sin \theta}}\partial_\theta (\sin\theta~\hat K^r_{~\theta}).
\label{momr}
\end{eqnarray}

Since the particles are confined to the equatorial plane $\theta=\pi/2$ where
$\beta^\theta = 0$, the particle 4-velocity component satisfies
$u^\theta=u_\theta=0$. Hence $S_\theta=0$.
Equation~(\ref{momtheta}) integrated across the equator yields
\begin{eqnarray}\label{intmomtheta}
\int_{-}^{+} r \sin \theta d \theta \left[
{1\over{\sin \theta}}\partial_\theta(\sin \theta~ \hat K^r_{~r})
\right.
\nonumber \\
+ {T\over{\sin^2\theta}}
\partial_\theta\left({{\sin^2\theta}\over T} \hat K^\phi_{~\phi}\right)
= -S_\theta
\nonumber \\
\left.
+ {1\over r^2}\partial_r (r^2 \hat K^r_{~\theta}) \right],
\end{eqnarray}
where $\pm$ denotes $\theta= \pi/2 \pm \epsilon,~~\epsilon \rightarrow 0$.
Functions that are symmetric across the equatorial plane, such as $\hat K^r_r$
and  $\hat K^\phi_\phi$, are continuous there.  Hence Eq.~(\ref{intmomtheta})
reduces to $0=0$.  Now consider Eq.~(\ref{momr}).  Integrating it
gives
\begin{equation}\label{intmomr}
0 = \int_{-}^{+} S_r r \sin \theta d \theta - {1 \over r} \hat K^r_\theta
|_-^+.
\end{equation}
Since $\hat K^r_\theta$ is antisymmetric across the equator, Eq.~\ref{intmomr}
gives
\begin{equation}\label{krtbound}
{\hat K}^r_{\theta} |^+ = - {\hat K}^r_{\theta} |^- = {r \over 2}
\int_-^+ S_r r \sin \theta d\theta .
\end{equation}
Similarly, integration of the Hamiltonian constraint equation (Eq.~(A6) of
Ref.~\cite{coll}) leads to
\begin{equation}\label{intham}
{1 \over r} \sin \theta \psi_{,\theta} |^+ = - {1 \over 4r} \psi \eta_{,
\theta}
- {1 \over 8 \psi}\int_-^+ \rho^* r \sin \theta d\theta .
\end{equation}
Integrating the lapse equation (Ref.~\cite{coll} Eq.~(A7)) gives
\begin{equation}\label{intlapse}
{1 \over r} \sin \theta (\alpha \psi)_{,\theta} |^+ = - {1 \over 4r} \alpha
\psi
\eta_{, \theta} |^+ \nonumber \\
+ {1 \over 8 }{\alpha \psi \over B} \int_-^+ (\rho^* + 2S) r \sin \theta
d\theta .
\end{equation}

The boundary condition Eq.~\ref{krtbound} is used to set the value of
$\hat K^r_\theta$  all along the equatorial plane.  In the vacuum, outside
of the equatorial plane, $\hat K^r_\theta$ is determined by integrating
the evolution equation,  Eq.~(A3) of Ref.~\cite{coll}, as usual.

When finite differencing the Hamiltonian constraint  (Eq.~A6 of
Ref.~\cite{coll}), the derivative terms $\psi_{,\theta}$ and $\eta_{,\theta}$
appear in exactly the combination as in Eq.~(\ref{intham}).  The only place
where the matter source term $\rho^*$ appears in the Hamiltonian
constraint is through this boundary condition.  Equation~\ref{intlapse}
is used in an analogous fashion for the lapse equation (Ref.~\cite{coll}
Eq.~A7).

The dynamical equation for $\eta$ (Ref.~\cite{coll} Eq.~A2) and
the shift equations (Ref.~\cite{coll} Eqs.~A8-A9) for
$\beta^r$ and $\beta^\phi$ remain unchanged.   Note that $\eta$,
$\beta^r$ and $\beta^\phi$ are metric coefficients and thus must
be continuous across the equator.

\subsection{Matter sources}

The geodesic equations of motion for the collisionless matter particles
are given by Eq.~(A10-A16) of Ref.~\cite{coll} with the following
simplifications: $u_\theta=0$ and  $\theta=\pi/2$.  Hence, as is
spherical symmetry, only the radial motion is dynamical for an
infinitely thin disk.

The particles are binned in annuli to determine the source terms
for the field equations.  Equations (A17-A21) of Ref.~\cite{coll}
lead to the disk sources~\cite{normnote}:
\begin{eqnarray}
\sigma \equiv \int_-^+ \rho^* r \sin \theta d \theta &=&
\sum_j {m \hat{u}_j \over (2 \pi r \Delta r)_j} \label{rhosum} \\
{\Sigma} _r \equiv \int_-^+ S_r  r \sin \theta d \theta &=&
\sum_j {m (u_r)_j \over (2 \pi r \Delta r)_j} \label{srsum} \\
{\Sigma} \equiv \int_-^+ S r \sin \theta d \theta &=&
\int_-^+ \rho^* r \sin \theta d \theta \nonumber \\ -
\sum_j {m \over {\hat u}_j (2 \pi r \Delta r)_j} \label{ssum}.
\end{eqnarray}
Here $m$ is the particle rest-mass related to the total rest-mass $M_0$
by $m=M_0/N$ with $N$ the total particle number.  We obtain $M_0$ from
\begin{equation}
\label{M0fromsigma0}
M_0=\int \sigma_0 2 \pi r dr,
\end{equation}
where
\begin{equation}
\sigma_0 \equiv \int_-^+ \rho_0 r \sin \theta d \theta =
\sum_j {m \over (2 \pi r \Delta r)_j} \label{rho0sum},
\end{equation}
and where $\rho_0$ is the rest-mass density.

\subsection{Hydrodynamical disks}\label{hydro}

In the special case in which concentric shells of particles do
not cross, collisionless matter may be treated as hydrodynamical
dust.  Thus, as an alternative to integrating geodesic equations
followed by binning of particles, one could integrate the equations of
relativistic hydrodynamics.  The hydrodynamical equations have the
disadvantage that they are PDE's and not ODE's like the geodesic
equations.  However, they have the advantage that they produce
intrinsically smooth source profiles, unlike particle
descriptions which are stochastic.

The basic equations of relativistic hydrodynamics in 3+1
ADM form are given in, e.g., Ref.~\cite{st80}.  For a cold
axisymmetric disk they reduce to the continuity and radial
Euler equations:
\begin{eqnarray}
\partial_t \sigma_0 + {1 \over r}
\partial_r(r \sigma_0 v^r) &=& 0, \label{cont} \\
\partial_t \Sigma_r + {1 \over r} \partial_r (r \Sigma_r v^r) &=&
- \sigma \partial_r \alpha + \Sigma_r \partial_r \beta^r  +
\Sigma \partial _r \ln A \alpha, \label{euler}
\end{eqnarray}
where
\begin{eqnarray}
\sigma&=&(\sigma_0^2 + {\Sigma_r^2 \over A^2})^{1/2}, \label{sigmadef} \\
v^r &=& {\alpha \over A^2} { \Sigma_r \over \sigma} - \beta^r, \label{vrdef} \\
\Sigma &=& { \Sigma _r^2 \over \sigma A^2} .
\label{Sigmadef}
\end{eqnarray}

\section{Analytic solutions and tests}\label{sec:analytic}

Before we consider numerical solutions of the dynamical equations
for disks and their gravitational fields,
we give here some analytic results that will serve as code
checks and initial data for our evolutions.

\subsection{Oscillating Newtonian disks}\label{newtdisk}
As discussed in Ref.~\cite{scalar}, there exists a complete analytic solution
that furnishes a good test of a numerical disk code in the weak field,
slow-motion limit. The solution
describes an oscillating homogeneous spheroid in Newtonian gravitation in
the disk limit (eccentricity $e \to 1$).
The surface density of such a  disk of mass $M$ and radius $R$ is
\begin{equation}\label{5.1}
\sigma(r)={3M\over2\pi R^2}\left( 1-{r^2\over R^2}\right)^{1/2}.
\end{equation}

Start with the equation of motion for the semi-major axis $R$ of an oblate
homogeneous spheroid (e.g. Eq. 5.6 of Ref.~\cite{st87}). Take the limit
$e\to 1$ and find
\begin{equation}\label{5.2}
\ddot R=-{3\pi\over 4}
{GM\over R^2} +{h^2\over R^3},
\end{equation}
where $h$ is the conserved angular momentum per unit mass of a particle at
the surface.
Since the motion is
homologous, the radius of each particle satisfies a similar equation.
Choose $h$ to be a fraction $\xi$ of the equilibrium angular momentum
$h_0=(3\pi GM R_0/4)^{1/2}$. Set
\begin{equation}\label{5.3}
R=R_0 X(t).
\end{equation}
Then the radius $r$ of each particle satisfies
\begin{equation}\label{5.4}
r = r_0 X(t),
\end{equation}
where $r_0$ is the initial radius. Substituting Eq.~(\ref{5.3}) into
Eq.~(\ref{5.2}) we see that $X$ satisfies the familiar
equation of an elliptic orbit for a particle with specific angular momentum
$h_{\rm eff}=\xi(M/R_0^3)^{1/2}$ around a fixed central mass
$M_{\rm eff}=3\pi M/4$.
The parametric solution for $X(t)$ is
\begin{eqnarray}
\label{5.5}
X &=& a (1-e \cos u),\\
t &=& {P \over 2 \pi} (u -e \sin u) - {P \over 2}.
\label{5.6}
\end{eqnarray}
where we assume $X=1$ and $\dot X=0$ at $t=0$.
In Eqs.~(\ref{5.5}) and (\ref{5.6}),
the semi-major axis, eccentricity and period are given by
\begin{eqnarray}
a &=& {1 \over 2 - \xi^2},\cr
e &=& 1 - \xi^2,\cr
P &=& 2 \pi \left( {4R_0^3 \over 3\pi GM (2- \xi^2)^3}
 \right)^{1 /2}.
\end{eqnarray}
The radial and tangential particle velocities are given by
\begin{eqnarray}\label{5.7}
v_r &=& {\dot X \over X} r, \\
v_{\phi} &=& \xi {r \over X^2} \left({3\pi GM \over 4 R_0^3}
\right)^{1/2}.
\end{eqnarray}

It is simple to derive the gravitational wave amplitude for an
oscillating disk in the quadrupole approximation.  In axisymmetry,
in the absence of rotation, there is only one polarization and its
amplitude is given by
\begin{equation}
\label{hplusgen}
r h_+ = {3 \over 2} \ddot I_{zz} \sin^2 \theta,
\end{equation}
where
\begin{equation}
\label{izz}
I_{zz} = - {1 \over 3} \int r^2 \rho d^3 x = - {2 \pi \over 3}
\int r^3 \sigma dr = - {2 \over 15} M R^2
\end{equation}
Here we have used Eq.~(\ref{5.1}).  Using
Eqs.(\ref{5.2}) and (\ref{5.3}) we find
\begin{equation}
\label{hplus}
r h_+ = -{2 \over 5} M \sin^2 \theta \left[ R_0^2 \dot X^2 + {3 \pi M \over 4
R_0}
\left( - {1 \over X} + {\xi ^2 \over X^2} \right) \right]_{t-r} .
\end{equation}

\subsection{Kalnajs Disk}\label{kaldisk}
When $\xi=1$, the above  Newtonian disk solution corresponds to a uniformly
rotating disk
in dynamical equilibrium.
As mentioned in the Introduction, such a disk is unstable
to the formation of rings but can be stabilized by heating.
Kalnajs~\cite{kalnajs}
has given an analytic prescription for constructing hot
homogeneous disks in equilibrium. They all have the same surface density
$\sigma$ and gravitational potential $\Phi$ as the cold disk in subsection
A, but differ in the amount of random motion. In these models, the
particles have an isotropic velocity distribution in a rotating frame that
moves with angular velocity $\Omega$:
\begin{eqnarray}
v \cos \chi &=& v_{\phi} - \Omega r, \\
v \sin \chi &=& v_r.
\end{eqnarray}

Here $v$ is the magnitude of the isotropic velocity in the rotating frame
and $\chi$ is a random angle about the particle position in the disk plane.
The distribution of particle velocities is given by
\begin{equation}
f (v) v dv = 2 \pi K \left[v_{{\rm max}}^2 - v^2\right]^{-{1 /2}} v dv.
\end{equation}
Here
\begin{eqnarray}
v_{{\rm max}}^2 &=& (\Omega_0^2 - \Omega^2) (R^2_0 - r^2), \\
\Omega_0 &=& {h_0 \over R^2_0},
\end{eqnarray}
and $K$ is a normalization constant.
Models in this family are parametrized by the ratio $\Omega/\Omega_0$.
Cold disks have $\Omega=\Omega_0$. Hot disks with $\Omega/\Omega_0 <
0.816$ are stable against ring formation.

\subsection{Relativistic disk}\label{reldisk}

Here we construct a relativistic generalization of the cold
homogeneous Newtonian disk described in Sec.~\ref{newtdisk}.
This is useful for studying the dynamical behavior of
disks in the strong field region, where almost nothing
is known.

Consider a disk of particles in circular equilibrium.
The Hamiltonian equation (Ref.~\cite{coll} Eq.(A6))
reduces to
\begin{equation}
\label{hamred}
\nabla^2 \psi = - 2 \pi {\rho^* \over \psi},
\end{equation}
where we have temporarily restored a factor of $8\pi$
to the right-hand-side.  We can obtain an analytic solution
if we first cast Eq.~(\ref{hamred}) into
Poisson's equation by setting  $\rho^*/\psi = 2 \rho_N$, obtaining
\begin{equation}
\label{hamred2}
\nabla^2 \psi = - 4 \pi {\rho_N},
\end{equation}
and then equating $\rho_N$ to the homogeneous Newtonian density profile
for an oblate spheroid in the pancake limit\cite{nst}.
We then find
\begin{equation}
\label{psifromphi}
\psi = 1 - \Phi_N,
\end{equation}
where $\Phi_N$ is the Newtonian potential for a flattened
spheroid~\cite{nst}.  In the disk interior, this
yields
\begin{equation}\label{phiinterior}
\Phi_N = - {3 \pi M_N \over 4 R_0} (1 - {1\over 2} {r^2 \over R_0^2})
{}~~~~~~~~~~
({\rm interior}) .
\end{equation}
The total mass of the disk $M$ is related to the Newtonian mass
$M_N$ appearing in $\Phi_N$ by
\begin{equation}
M=2M_N.
\end{equation}

Comparing Eqs.~(\ref{hamred}) and (\ref{hamred2})
we find that the surface density Eq.~(\ref{rhosum})
is given by
\begin{equation}
\sigma= 2\psi \sigma_N,
\end{equation}
where $\sigma_N$ is the corresponding Newtonian density
given by Eq.~(\ref{5.1}).

The angular motion of a particle in an equilibrium disk
is determined by setting $du_r/dt = u_r=0$ and
using Eqs.~(A.8) and (A.13) of Ref.~\cite{coll}.
The result is
\begin{equation}
\label{uphicalc}
u_\phi^2 = {(\alpha_{,r}/\alpha) B^2 r^2 \over
B_{,r}/B - {\alpha_{,r}/\alpha + {1/r}}} ~~~~~~~~~~
({\rm equilibrium}),
\end{equation}
where $B= \psi^2$ is given by
Eqs.~(\ref{psifromphi}) and (\ref{phiinterior}).
The lapse $\alpha$ is found from the maximal slicing ($K^i_i=\partial_t
K^i_i=0$)
condition, Eq.~(A7) of Ref.~\cite{coll} specialized to equilibrium,
for which $\eta=0=\hat{K}^i_j$.
The source term $\Sigma$ appearing in the jump condition for
the lapse equation is given by
\begin{equation}
\Sigma = \sigma { u_\phi ^2 \over B^2 r^2 + u_\phi ^2}
\end{equation}
where we have used Eq.~(A16) of Ref.~\cite{coll}.
Because $\alpha$ and $u_\phi^2$ are interdependent, it
is necessary to iterate the lapse equation and Eq.(\ref{uphicalc}).
As an initial guess we use $\alpha \simeq 1 + \Phi_N^{(exterior)}$.

The rest-mass surface density $\sigma_0$ defined in
Eq.(\ref{rho0sum}) is given by
\begin{equation}
\sigma_0= {2 \sigma_N \psi \over (1 + u_{\phi}^2 /B^2 r^2)^{1/2}}
\end{equation}
from which the total rest mass can be computed using Eq.~(\ref{M0fromsigma0}).

We can obtain a solution to the initial value equations for a
{\it nonequilibrium} disk at a moment of time symmetry when the
particles are moving at a fraction $\xi$ of their equilibrium
velocity:
\begin{equation}
u_\phi = \xi u_\phi^{(equil)} .
\end{equation}
The other equations remain unchanged.  In the Newtonian limit,
this solution goes over to the cold oscillating disk of Sec.~\ref{newtdisk}.

An interesting nonequilibrium solution is the one in which $\xi=0$.
This corresponds to the collapse of a homogeneous disk in which all
the particles are at rest initially.  This is the pancake analogue of
Oppenheimer-Snyder collapse of a homogeneous sphere.  In the disk
case, however,  the solution is radiative and is not known analytically.
In this case the total mass and rest mass are related by
\begin{equation}
M = {2 M_0 \over 1 + (1 + 6 \pi M_0/5 R_0)^{1/2} },
\end{equation}
(see Ref.~\cite{nst}).
\section{Numerical results}\label{sec:results}
In this section, we give examples of evolutions
computed with our fully relativistic particle disk plus
gravity code.  When possible, we show gravitational
wave forms and also make contact with analytic results.
We also monitor quasi-local mass indicators; for the runs
shown here the total mass is numerically preserved to within
about $3\%$.  In Table I we summarize the cases discussed
in the text.
\subsection{Oscillating Cold Newtonian disk}

As a check on our code, one would first
like to simulate the collapse of the oscillating
homogeneous disk described in Sec. \ref{newtdisk}.
Unfortunately, such a cold
configuration is unstable to ring formation.
This is the familiar result for
cold equilibrium disks with $\Omega=\Omega_0$ and
$\xi=1$ discussed above; we find numerically
that it also holds for cold oscillating disks.
We can still employ this analytic model
to check the field solver in our code provided we
supply the unperturbed source function $\sigma$ analytically as a function
of time. We then allow the code to solve the field equations and
thereby determine the gravitational radiation numerically
from the metric and extrinsic curvature.

\begin{figure}
\special{hscale=0.45 vscale = 0.45 hoffset = -0.2 voffset = -4.25
psfile=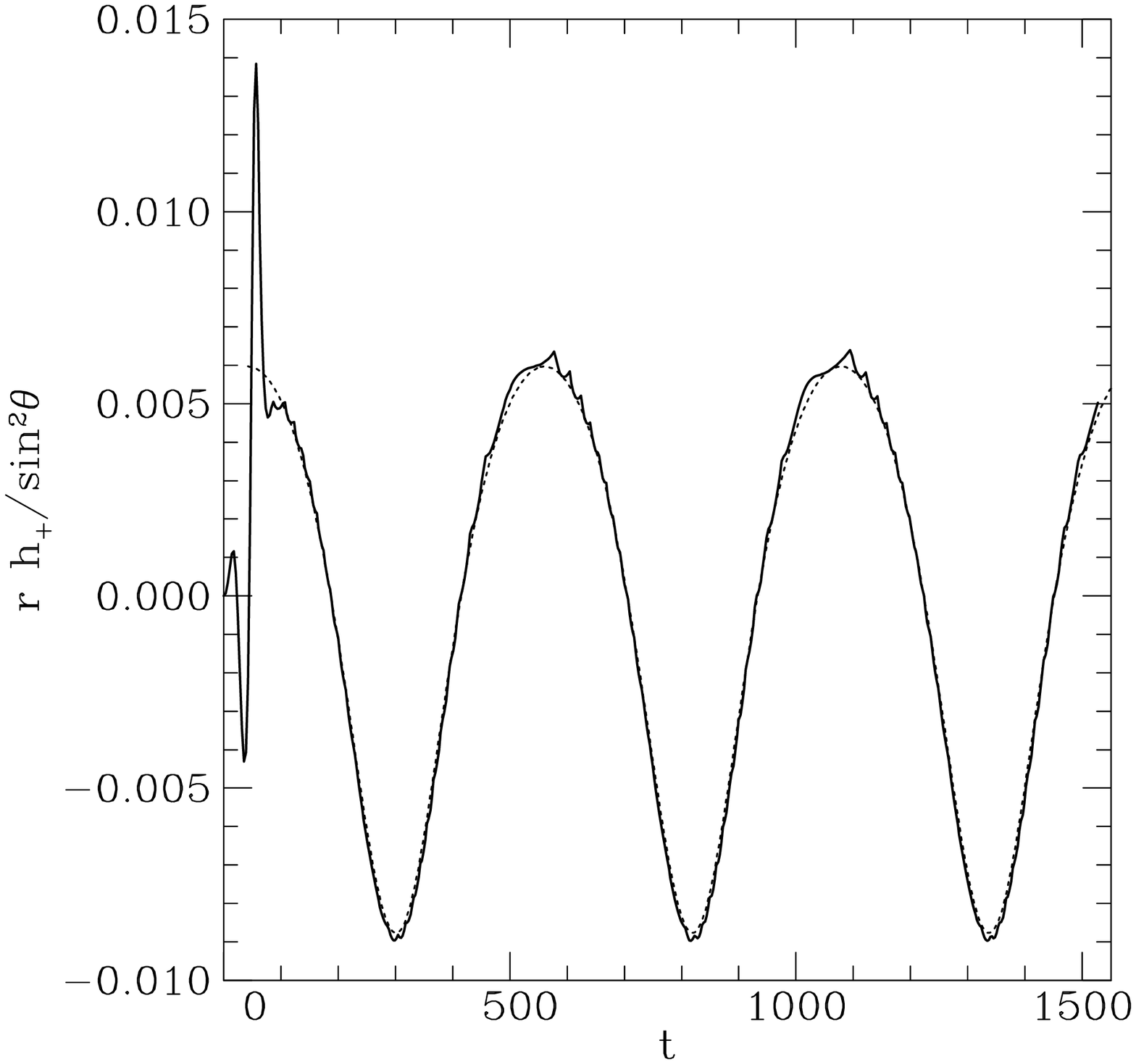}
\vspace{3.5in}
\caption{Gravitational wave form from a cold oscillating
disk with an analytic Newtonian matter source.
The numerically generated and propagated
gravitational wave form (solid line), extracted at a radius
of $41M$, is compared with the analytic prediction,
Eq.~(30)(dotted line).
The disk has an equilibrium radius
$R_0/M=30$ and a velocity cutdown $\xi=0.90$.  The wave
amplitude $h_+$ is dimensionless while $r$ and $t$
are in units of $M$.}
\label{fig:coldndisk}
\end{figure}

Here we give results from a typical evolution of a
homogeneous disk with $R_0/M_0=30$ and
velocity cutdown factor $\xi=0.9$.
This choice of radius is sufficiently large that the system remains
essentially Newtonian throughout its evolution.
We place the outer boundary at $r_{\rm max}/M_0=500$ and
use a mesh with 30 radial zones inside the
matter, 220 outside, and 16 angular zones in the
upper hemisphere.
In Fig.~\ref{fig:coldndisk}
we compare the gravitational
wave form extracted at an arbitrary fixed
exterior radius of $42M$ using a
gauge-invariant technique~\cite{ae90}, with the
analytic prediction Eq.~(\ref{hplus}).  After
an initial transient, the wave forms coincide very
closely.  The slight noise in the extracted wave form
arises from interpolation error when the analytic
source is mapped onto the numerical grid.

The excellent agreement between the analytic and numerical
results here gives confidence in the reliability of
the numerical implementation of the jump conditions
and the solution of the field equations.

\subsection{Oscillating Kalnajs disk}

To test the particle simulation aspects of our code,
we set up stable equilibrium Kalnajs disks
as described in Sec.~\ref{kaldisk}. Our code successfully holds
these in equilibrium for several rotation periods.

\begin{figure}
\special{hscale=0.45 vscale = 0.45 hoffset = -0.45 voffset = -4.
psfile=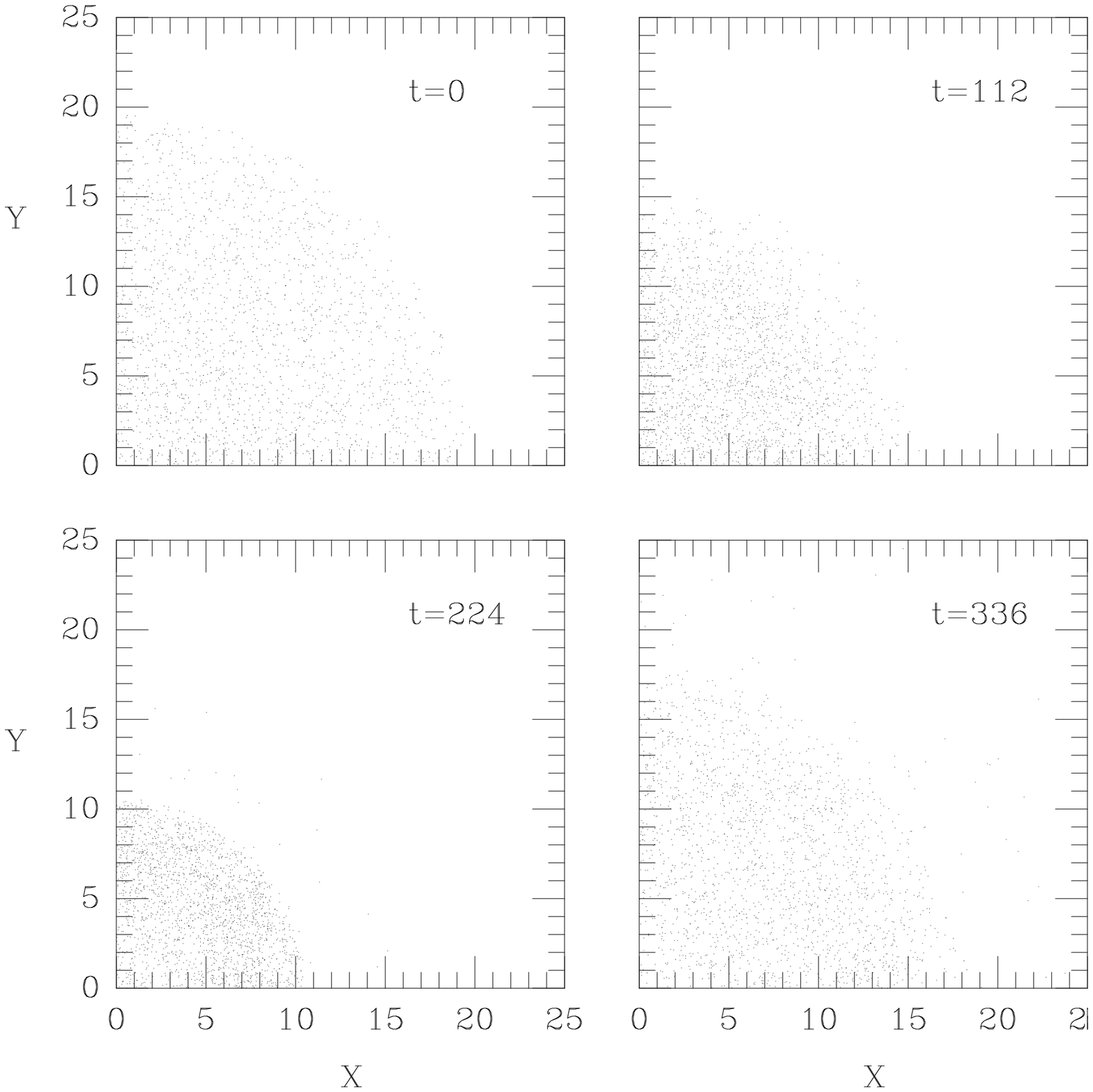}
\vspace{3.5in}
\caption{Snapshots of the particle positions for
the evolution of an oscillating Kalnajs disk.  The
initial radius is $R_0/M=20.0$,
velocity cutdown $\xi=0.9$, and velocity dispersion
parameter $\Omega/\Omega_0=0.8$.  Particle coordinates
and time are in units of $M$.}
\label{fig:kalpart}
\end{figure}

We next study the production of gravitational radiation
from the oscillation of a hot nonequilibrium  disk.
We set up a Kalnajs disk with $R_0/M_0=20$ and
velocity dispersion $\Omega/\Omega_0=0.8$.
We then induce collapse by reducing all particle velocity components by
a cutdown factor $\xi=0.9$.  This run employed $200$
radial by $32$ angular zones and
12000 particles. Frames from the evolution are shown
in Fig.~\ref{fig:kalpart}.

\begin{figure}
\special{hscale=0.45 vscale = 0.45 hoffset = -0.2 voffset = -4.25
psfile=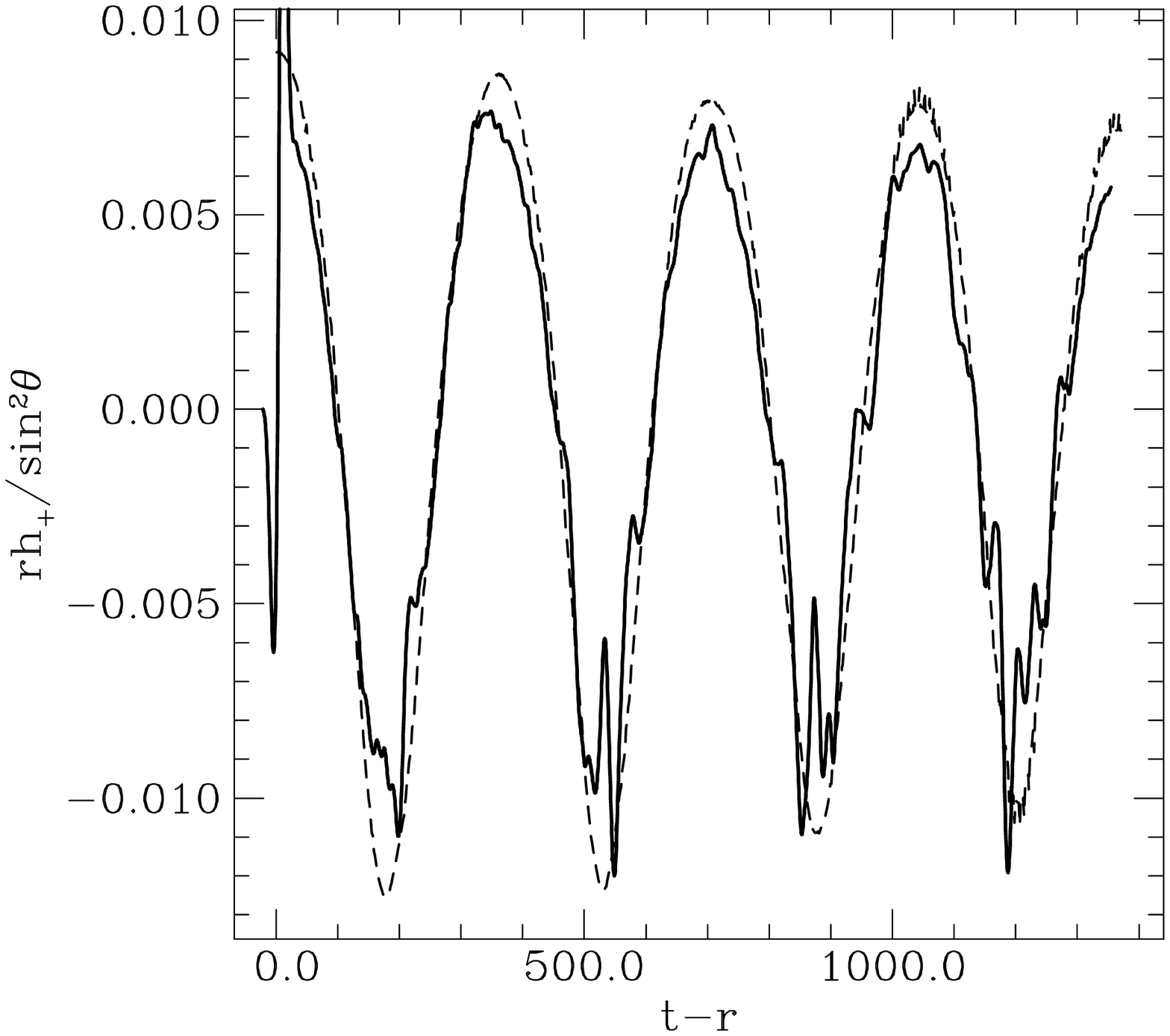}
\vspace{3.5in}
\caption{Gravitational wave form from the oscillating Kalnajs disk
shown in Fig.~2. The quadrupole wave form extracted
at a radius of $r/M=21$ (solid line)
is compared with the quadrupole formula result, Eq.~(28)
(dashed line).  }
\label{fig:kalwave}
\end{figure}

Although some of the collisionless
matter from the hot disk escapes, most of the particles stay together
and oscillate homologously around the equilibrium radius.
In Fig.~\ref{fig:kalwave}
we compare the extracted quadrupole wave form with $h_+$
computed numerically from the particle positions
and velocities according to the quadrupole formula Eq.~(\ref{hplusgen})
with
\begin{equation}
I_{zz} = - {1 \over 3} \sum _j m r_j^2.
\end{equation}
As this disk is nearly Newtonian,
higher multipoles are expected contribute negligibly to $h_+$.
The good agreement over several oscillations in Fig.~\ref{fig:kalwave} tests
many
facets of the mean-field
particle simulation scheme, including the field and particle integrators,
the particle and grid sampling, and the source binning algorithm.
Wave forms extracted at different radii agree at the $15\%$ level
at this grid resolution. Our experiments indicate
that most of this difference, and the
difference from the quadrupole calculation, arises from instantaneously
propagated errors in solutions of the constraint equations.
To remove near-zone and gauge effects, the wave extraction method requires
delicate cancellations between different multipole moments.
Counter-streaming in the particle source can cause inconsistencies
between the hyperbolically and elliptically computed parts of
the gravitational field, making this cancellation inaccurate.

\subsection{Cold equilibrium relativistic disk}

\begin{figure}
\special{hscale=0.45 vscale = 0.45 hoffset = -0.45 voffset = -4.
psfile=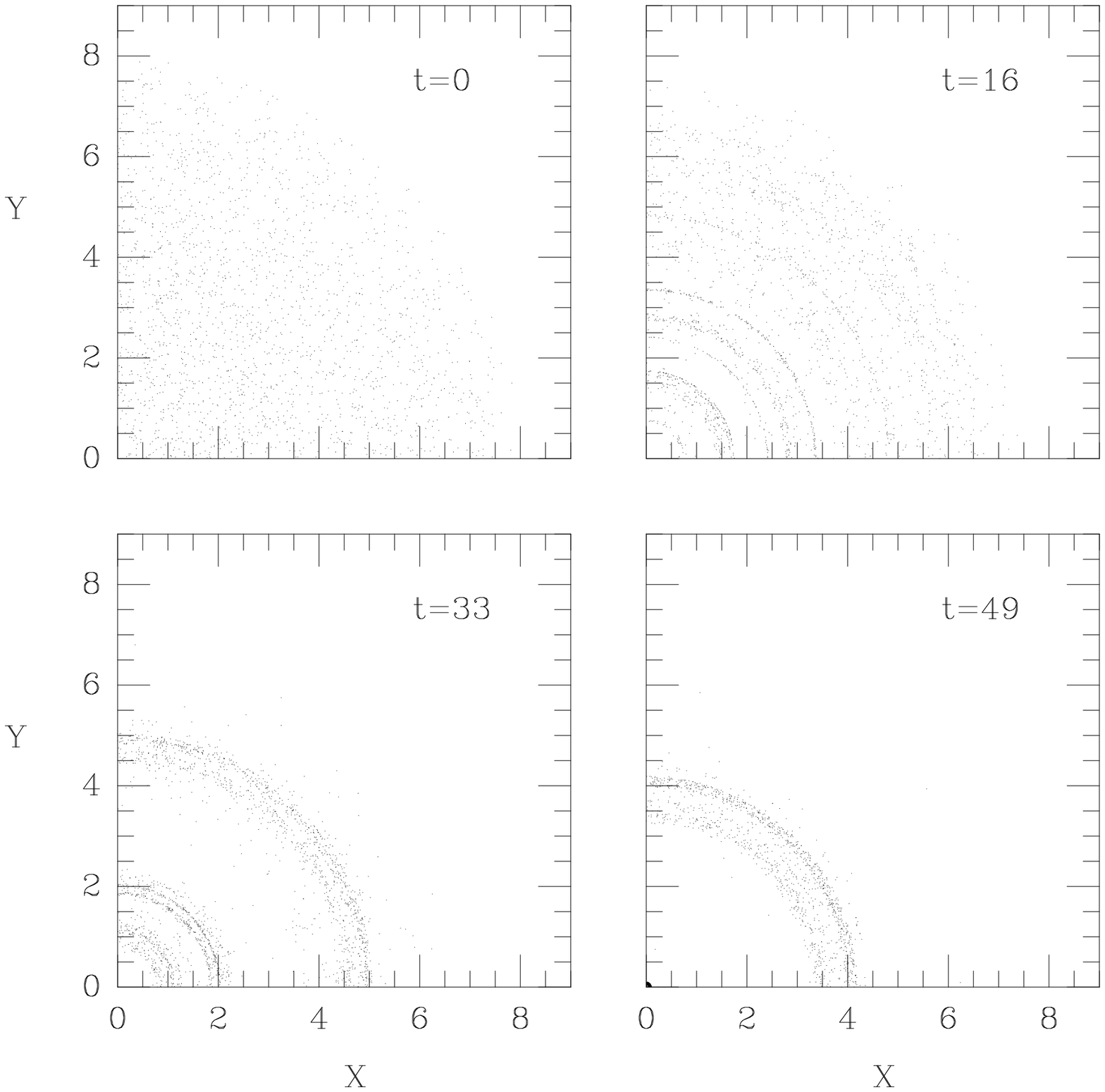}
\vspace{3.5in}
\caption{Snapshots of the particle positions for the evolution
of a cold, equilibrium relativistic disk.
The initial disk radius is  $R_0/M=8.0$.
The rapid growth of concentric rings is apparent.  }
\label{fig:coldrdisk}
\end{figure}

Here we consider equilibrium disks constructed
according to the prescription of Sec.~\ref{reldisk}.  The question
we wish to answer is whether this disk is unstable to
rings as in the Newtonian theory when
the source and gravitational field are evolved
consistently.  In Fig.~\ref{fig:coldrdisk}
we show a relativistic disk with $R_0/M=8$.
Before the evolution progresses very
far, ring formation begins and by $t=30M$ is completely
dominant.  Less relativistic configurations behave similarly.
This calculation was carried out with $300$ radial by $16$ angular
zones and 12000 particles, but runs with as many
as 48000 particles did not not slow down the ring formation.
\subsection{Hot relativistic disk}
\begin{figure}
\special{hscale=0.45 vscale = 0.45 hoffset = -0.45 voffset = -3.75
psfile=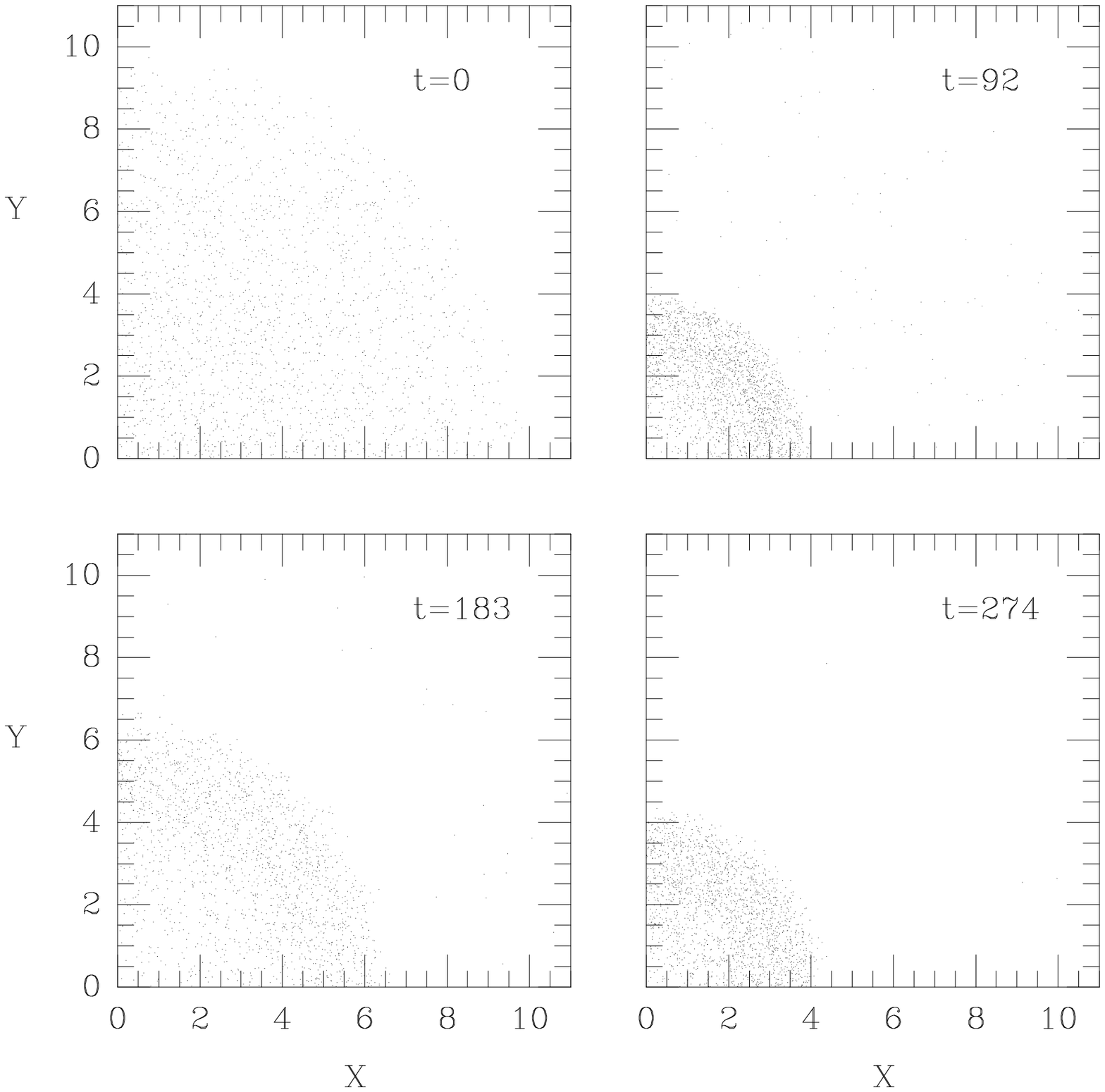}
\vspace{3.0in}
\caption{ Snapshots of the particle positions for the evolution of a
hot, near-equilbrium relativistic disk.  The initial disk radius is
$R_0/M=10.0$ and the velocity dispersion parameter $\Omega/\Omega_0 = 0.85$.
Following collapse, the disk settles down to a new equilbrium state.}
\label{fig:hotrdiskp}
\end{figure}

As discussed before, ring formation is suppressed by the
addition of sufficient random particle motion.  Unfortunately,
only small values of velocity dispersion can be used without
dissipating the outer region of the disk.  Here we consider the evolution
of a relativistic disk with $R_0/M=10$, and velocity dispersion
parameter $\Omega/\Omega_0=0.85$ and $\xi=1.0$.
Because of its compactness, this near-equilibrium relativistic Kalnajs
disk is unstable to collapse.
As shown in Fig.~\ref{fig:hotrdiskp}
the disk
initially collapses and then oscillates about
a new equilibrium of about $R_0/M=6.0$.   At late times it virializes
to a static equilibrium state.

\begin{figure}
\special{hscale=0.45 vscale = 0.45 hoffset = -0.2 voffset = -4.5
psfile=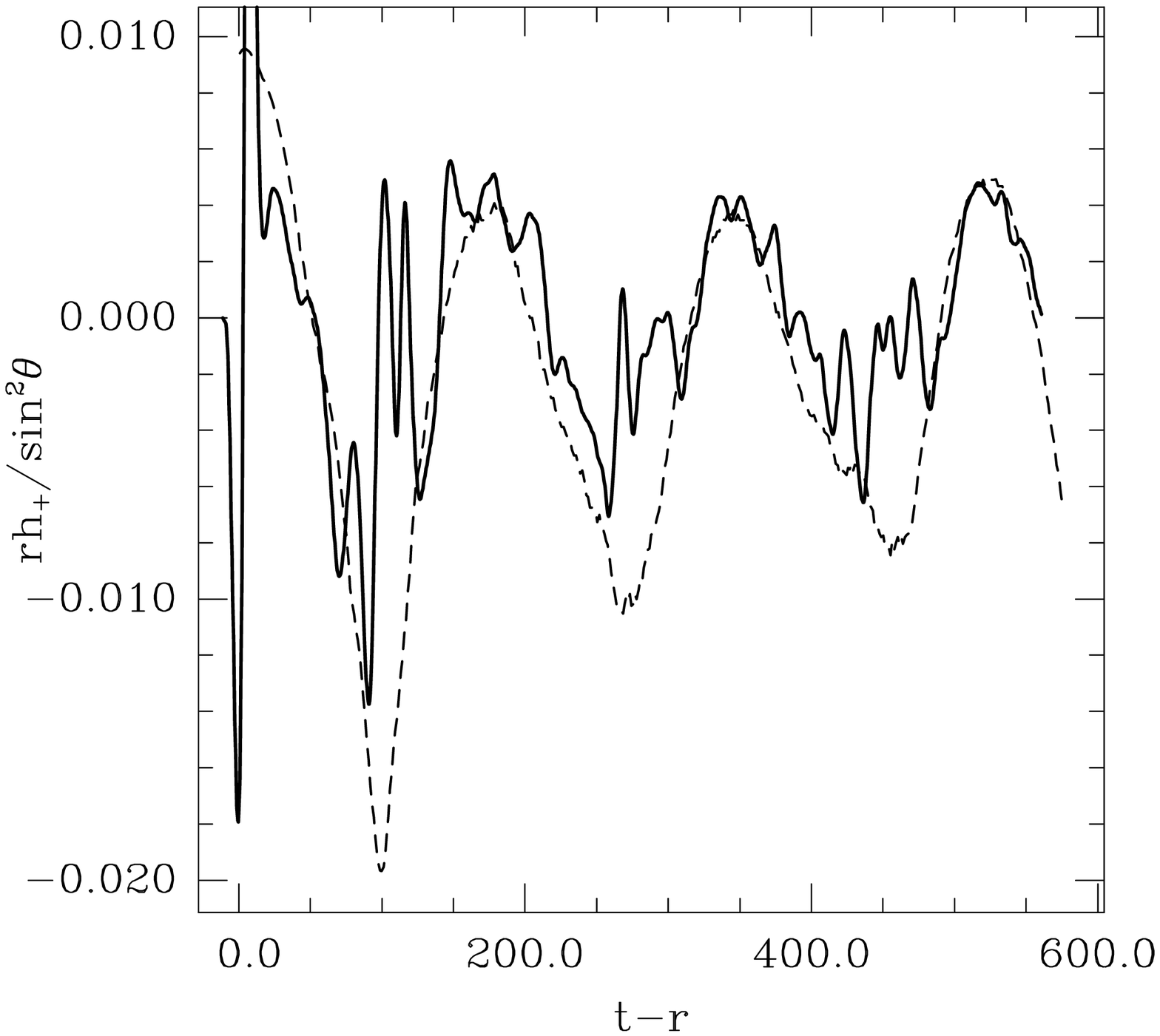}
\vspace{3.5in}
\caption{Wave form from the hot relativistic disk shown in Fig.~5.
Wave amplitudes are labeled as in Fig.~3.  }
\label{fig:hotrdiskw}
\end{figure}

In Fig.~\ref{fig:hotrdiskw}
the gravitational wave form extracted at $r/M=12.0$ is
compared with the quadrupole formula result.  As discussed above,
we have evidence that the stochasticity of the particle
source is responsible for the short-time scale discrepancies
between the wave forms.  This calculation used a $200$ radial
by $16$ angular zones and 12000 particles.  Improving
the accuracy requires higher grid resolution and more particles;
unfortunately the noise in the wave form is a $\sqrt N$ effect
so increasing the number of particles quickly becomes
computationally prohibitive.
\subsection{Cold disk collapse}
\begin{figure}
\special{hscale=0.45 vscale = 0.45 hoffset = -0.45 voffset = -3.75
psfile=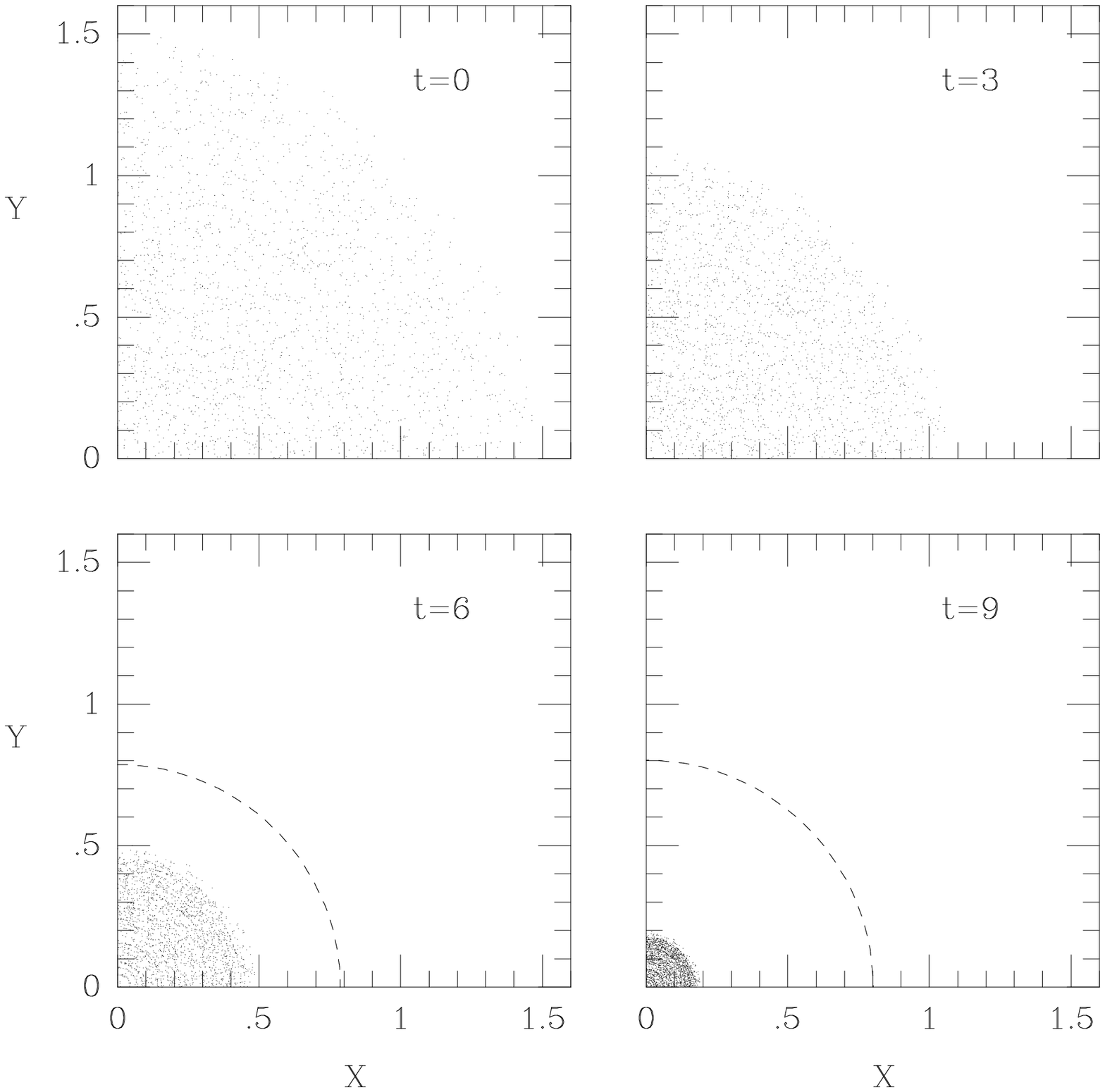}
\vspace{3.in}
\caption{Snapshots of the particle positions for
the collapse of a cold relativistic disk.  Initially the radius is
$R_0/M=1.5$ and the particles are all at rest.
The apparent horizon (dashed line) first appears
at $t \simeq 4.0M$.  }
\label{fig:ospart}
\end{figure}

Finally, we consider the collapse of a cold homogeneous disk,
the disk analogue of Oppenheimer-Snyder collapse in
spherical symmetry.
Initially, all particle velocities are equal to zero ($\xi=0$).
We consider a very relativistic case with $R_0/M=1.5$.
(Recall that in isotropic coordinates a Schwarzschild
black hole has a radius $R_0/M=0.5$.)
With such a compact disk, the collapse is quick enough
that an apparent horizon appears before the ring
instability becomes significant.  In Fig.~\ref{fig:ospart}
we show results from an evolution carried out
with $300$ radial by $16$ angular zones and
24000 particles.  The apparent horizon appears at a
time of about $4.0M$ and ring formation is just discernible
at the disk center at this time.

In order to prolong the numerical evolution after the black hole
forms and counteract the effect of ``throat stretching''
that occurs in our maximal (and other singularity avoiding)
time-slicing conditions, we have implemented a moving
mesh algorithm that moves the inner radial grid-zone to
track the growth of the conformal factor $\psi$.  With
this method, we are able to evolve the black hole and preserve
constancy of out quasi-local mass measures (the Brill and
ADM masses) to a few percent for around $15M$ after hole
formation.  Evolution beyond this time is prevented by
the numerical difficulties in integrating the particle
geodesics on the extremely stretched mesh.

\begin{figure}
\special{hscale=0.45 vscale = 0.45 hoffset = -0.2 voffset = -4.
psfile=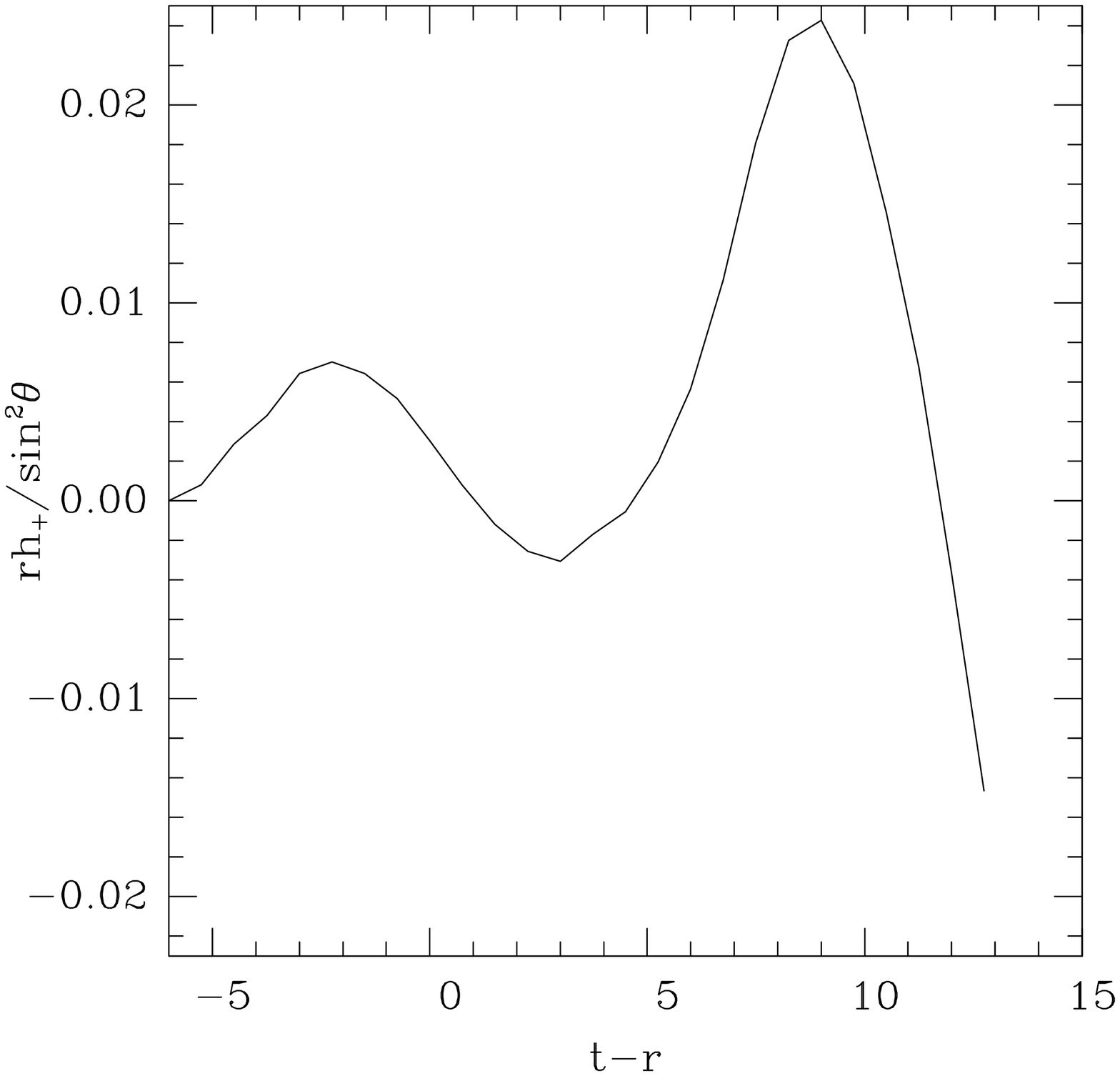}
\vspace{3.5in}
\caption{Gravitational wave form from the disk collapse
shown in Fig.~7.  The
quadrupole wave form extracted at $r/M=6.0$ is
shown as a function of retarded time. }
\label{fig:oswave}
\end{figure}

This time is sufficient, however, to see the pulse of
radiation from collapse and, perhaps, the first half wavelength
of a quasi-normal mode oscillation.  In Fig.~\ref{fig:oswave}
we show an estimate of the asymptotic quadrupole wave form
extracted at $r/M=6.0$.  After an initial peak representing
radiation in the initial data we see an oscillating signal
with total wavelength of approximately  $16M$,
comparable with that of the most
slowly damped $\ell=2$ quasi-normal mode
oscillation  which has $\lambda=16.8M$.
The energy radiated is less than $0.1\%$ in
the quadrupole mode and we estimate
that it is less than $0.01\%$ in higher $\ell$-modes.

\begin{figure}
\special{hscale=0.45 vscale = 0.45 hoffset = -0.2 voffset = -4.10
psfile=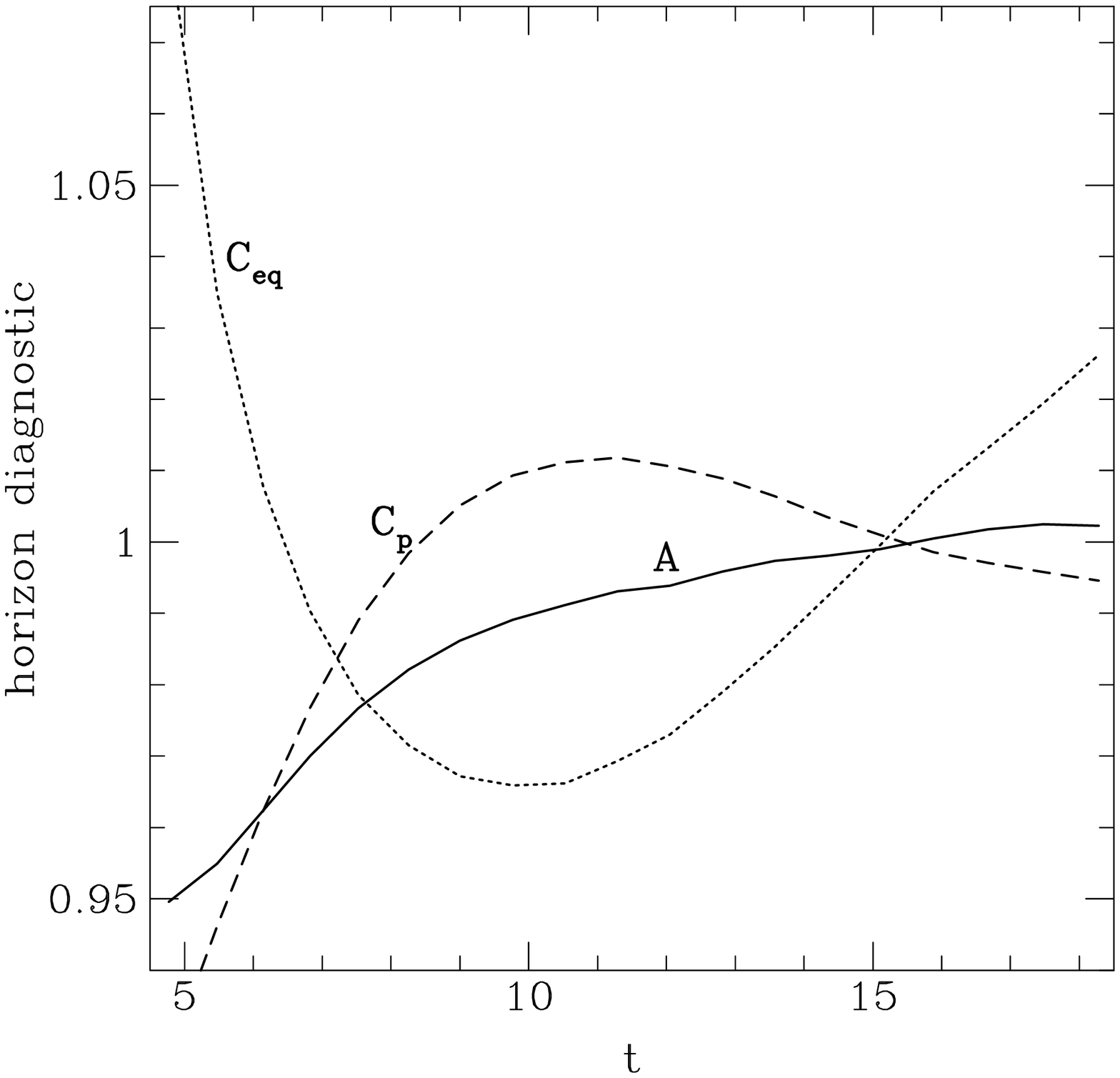}
\vspace{3.5in}
\caption{Horizon diagnostics for the black hole
formation shown in Fig. 7.  The proper
equatorial circumference, polar circumference,
and area of the apparent horizon normalized
to their Schwarzschild values,
${\cal C}_{eq}/4\pi M$,
${\cal C}_{p}/4\pi M$ and,
${\cal A}/16 \pi M^2$ are plotted as functions
of time in units of $M$.
}
\label{fig:oshor}
\end{figure}

We monitor the area of the apparent
horizon and its polar and equatorial circumferences
(see Ref.~\cite{coll} for definitions).
As shown in Fig.~\ref{fig:oshor},
the normalized apparent horizon area ${\cal A}/16 \pi M^2$
has a value of $0.95$ when it first forms, and asymptotes towards
the Schwarzschild value of $1.0$ before the
calculation terminates.   The black hole is
initially oblate when the apparent horizon
forms. The normalized proper circumferences
oscillate around their Schwarzschild
values ${\cal C}_{eq}/4 \pi M={\cal C}_p/4 \pi M = 1.0$
as the black hole radiates off its initial asphericity.
This oscillation period again is comparable with
that of the most slowly damped $\ell=2$ quasi-normal
mode.

\section{Conclusions}\label{sec:conclusions}
We have presented here preliminary results from
a new relativistic mean-field,
particle simulation code that can evolve disks of collisionless matter
and compute the emitted gravitational radiation.  The
disk analogue to the Oppenheimer-Snyder solution,
cold homogeneous disk collapse, is an interesting benchmark for numerical
relativity codes as it is one of the simplest scenarios
that encounters the two most computationally challenging
features of relativistic collapse:  black hole formation and
evolution, and gravitational wave production and
propagation.
The hydrodynamical formulation of this problem, provided
in Sec.~\ref{hydro}, should permit implementation of disk
collapse
in codes without collisionless matter sources.

Failure to complete the computation of the full
radiation data in the case of disk collapse to
a black hole reflects the fundamental problem in
numerical relativity.  Specifically, no general
algorithm is currently known that can integrate
Einstein's equations for long enough
after a black hole forms to compute the full
gravitational wave form.  It may be possible
that this problem can be solved with
suitable apparent horizon boundary conditions, "cutting out"
the black hole (see e.g. Ref~\cite{seidelsuen}).
We expect disk collapse to be a useful proving ground
for developing apparent-horizon boundary conditions
and other black hole evolution techniques in an axisymmetric context.

\section*{Acknowledgements}
This work was supported by National Science Foundation
grants AST 91-19475 and PHY 90-07834 and by
the Grand Challenge grant NSF PHY 93-18152 / ASC 93-18152
(ARPA supplemented).  Computations
were performed at the Cornell Center for Theory and
Simulation in Science and Engineering, which is supported
in part by the National Science Foundation, IBM Corporation,
New York State, and the Cornell Research Institute.
%

%
%
%
%
%

%
%
%
\twocolumn
\widetext
\begin{table}
\caption{Disk Evolution Cases}
\label{table}
\begin{tabular}{l|c|c|c|c}
Case&$R_0/M$&$\Omega/\Omega_0$&$\xi$&fate\\
\hline
Oscillating cold Newtonian (analytic source) &30.0&1.0&0.9& oscillates \\
Oscillating Kalnajs&20.0&0.8&0.9& oscillates \\
Cold relativistic equilibrium&8.0&1.0&0.99& rings \\
Hot relativistic near-equilibrium&10.0&0.85&1.0& collapses and virializes\\
Cold relativistic&1.5&1.0&0.0& collapses to black hole
\end{tabular}
\end{table}


\begin{references}
\bibitem{normnote} Note that the source terms appearing in the appendix
of Ref.~\cite{coll} and used here are related to the components of
\protect{$T_{\mu \nu}$} by a factor \protect{$8 \pi A^2 B$}.
Hence, \protect{$\rho^*= 8 \pi A^2 B T_{\mu \nu} n^\mu n^{\nu}$} etc.
\end{references}
\end{document}